\documentclass{PoS}

\title{New implementation of the sector decomposition on FORM}

\ShortTitle{New implementation of the sector decomposition on FORM}

\author{\speaker{Takahiro Ueda} and Junpei Fujimoto \\
        High Energy Accelerator Research Organization (KEK), \\
        1-1 Oho, Tsukuba, Ibaraki 305-0801, Japan. \\
        E-mail: \email{uedat@post.kek.jp},
                \email{junpei@post.kek.jp}}

\abstract{%
Nowadays the sector decomposition technique, which can isolate divergences
from parametric representations of integrals, becomes a quite useful tool
for numerical evaluations of the Feynman loop integrals.
It is used to verify the analytical results of multi-loop integrals
in the Euclidean region, or in some cases practically used in the physical
region by combining with other methods handling the threshold.
In an intermediate stage of the sector decomposition for the multi-loop
integrals, one often has to handle enormously large expressions containing
tons of terms. The symbolic manipulation system {\tt FORM} is originally
designed to treat such huge expressions and has a strong advantage for it.
In this talk, the implementation of the sector decomposition algorithm
on {\tt FORM} is discussed. A number of concrete examples including cases
of multi-loop diagrams are also shown.
}

\FullConference{XII Advanced Computing and Analysis Techniques in Physics Research\\
  November 3-7, 2008\\
  Erice, Italy}

\usepackage{bm}
\usepackage{amsmath}
\usepackage{graphicx}

\makeatletter

\def\cal{\mathcal}
\def\sl#1{\mathpalette\@sl{#1}}
\def\@sl#1#2{\ooalign{\hfill$#1/$\hfill\crcr$#1#2$}}
\def\bar#1{\mathpalette\@bar{#1}}
\def\@bar#1#2{%
  \bgroup
    \settowidth\@tempdima{$#1#2$}%
    \@tempdima.8\@tempdima
    \overset{\underline{\hskip\@tempdima}}{#2}
  \egroup
}

\def\<{\langle}
\def\>{\rangle}
\def\l#1{%
  \ifnum 0<0#1 \def\@tempa{\@l} \else \def\@tempa{\left} \fi \@tempa#1%
}
\def\r#1{%
  \ifnum 0<0#1 \def\@tempa{\@r} \else \def\@tempa{\right} \fi \@tempa#1%
}
\def\@l#1#2{%
  \ifcase #1 #2 \or \bigl#2 \or \Bigl#2 \or \biggl#2 \or \Biggl#2%
    \or \bigggl#2 \or \Bigggl#2 \or \biggggl#2 \or \Biggggl#2 \else #2%
  \fi
}
\def\@r#1#2{%
  \ifcase #1 #2 \or \bigr#2 \or \Bigr#2 \or \biggr#2 \or \Biggr#2%
    \or \bigggr#2 \or \Bigggr#2 \or \biggggr#2 \or \Biggggr#2 \else #2%
  \fi
}
\def\bBigg@#1#2{%
  {\@mathmeasure\z@{\nulldelimiterspace\z@}%
     {\left#2\vcenter to#1\big@size{}\right.}%
   \box\z@}}
\def\biggg{\bBigg@3} 
\def\Biggg{\bBigg@{3.5}} 
\def\bigggg{\bBigg@4} 
\def\Bigggg{\bBigg@{4.5}} 
\def\bigggl{\mathopen\biggg}
\def\bigggr{\mathclose\biggg}
\def\Bigggl{\mathopen\Biggg}
\def\Bigggr{\mathclose\Biggg}
\def\biggggl{\mathopen\bigggg}
\def\biggggr{\mathclose\bigggg}
\def\Biggggl{\mathopen\Bigggg}
\def\Biggggr{\mathclose\Bigggg}

\def\ep{\epsilon}

\makeatother

\begin{document}

\section{Introduction}

The Large Hadron Collider (LHC) has started its operation,
and International Linear Collider (ILC) is planned as a next generation
collider. In order to obtain as much information as possible from such
experiments, it is undoubtedly important to know the precise predictions
from the theory at high energies. This fact pushes forward recent study
on the computation of one- and higher loop corrections in the perturbation
theory.

To evaluate the loop integrals, one may try to find formulae of them
analytically. However, if they have many loop momenta, many external legs,
and many kinematic parameters, it becomes an extremely difficult problem.
Another approach to the evaluation of the loop integrals is to perform
loop integrals by using a numerical integration method.
In principle, it is possible to construct a general recipe for computing
loop integrals numerically, which is applicable to a wide enough class of
the loop integrals. Even if one has an analytical formula for some integral,
it is good idea to have another method for computing any integrals numerically,
for checking purpose. In practice, to perform loop integrals numerically,
one has to handle several singularities: UV, IR divergences, and
singularities from physical thresholds.

In this talk, the implementation of the sector decomposition
algorithm~\cite{Binoth:2000ps,Binoth:2003ak}%
\footnote{%
  See~\cite{Heinrich:2008si} for a review.
  Currently, available public codes of the sector decomposition
  are~\cite{Bogner:2007cr,Smirnov:2008py}.
}%
which can isolate divergences from parametric representation
(e.g., Feynman parameters) of integrals, on a symbolic manipulation system
{\tt FORM}~\cite{vermaseren-2000}, is discussed.
Then, an application of the numerical extrapolation method on handling
physical thresholds~\cite{deDoncker:2004fb} after isolating all IR divergences
is also discussed.

\section{The sector decomposition for IR divergent loop integrals}

Here we briefly sketch how the sector decomposition algorithm is applied
for IR divergent loop integrals, in a easy example.
We consider the one-loop scalar massless on-shell box diagram
in $D=4-2\ep$ dimensions.
By introducing Feynman parameters and performing the momentum integration,
we arrive to
\begin{equation}
  \begin{split}
    I_4^{0m}
      &= \int\frac{d^Dk}{i\pi^{D/2}}
         \frac{1}{(k^2+i0)\l1[(k+p_1)^2+i0\r1]\l1[(k+p_{12})^2+i0\r1]\l1[(k+p_{123})^2+i0\r1]} \\
      &= \Gamma(2+\ep) \int_0^1 d^4x \, \delta(1-x_{1234})\frac{1}{(-s x_1x_3-t x_2 x_4 -i0)^{2+\ep}},
  \end{split}
  \label{eq:box}
\end{equation}
where $s=(p_1+p_2)^2$, $t=(p_2+p_3)^2$, and $p_i^2=0$ $(i=1,\ldots,4)$,
and we have used abbreviations, $p_{12}=p_1+p_2$, $p_{123}=p_1+p_2+p_3$
and $x_{1234}=x_1+x_2+x_3+x_4$. This integral has overwrap singularities
on edges of the integration region: the denominator of the integrand
vanishes when two variables simultaneously go to zero, e.g., $x_1,x_2 \to 0$.
The sector decomposition algorithm can disentangle such overwrap singularities,
by splitting integral domain and making variable replacements, iteratively.
One can obtain
{
\allowdisplaybreaks
\begin{align}
  I_4^{0m}/\Gamma(2+\ep)
    &= 2 \int_0^1 d^3x \l4[
         x_1^{-1-\ep} x_2^{-1-\ep} \frac{(1+x_{12}+x_1x_2x_3)^{2\ep}+(1+x_1+x_2x_{13})^{2\ep}}{(-s-tx_3-i0)^{2+\ep}} \notag \\
    &\hspace{80pt}
         + x_1^{-1-\ep} \frac{(1+x_{12}+x_1x_3)^{2\ep}}{(-s-tx_2x_3-i0)^{2+\ep}}
       \r4] \ +\  (s\leftrightarrow t).
\end{align}
}
Note that, in this form, there are no overwrap singularities,
and all singularities are factorised as powers of monomials. This can be
expanded with respect to $\ep$ as:
\begin{equation}
  I_4^{0m}/\Gamma(2+\ep) = \frac{C_2}{\ep^2} + \frac{C_1}{\ep} + C_0 + \cal{O}(\ep) ,
  \label{eq:box-integrand}
\end{equation}
with
\begin{subequations}
  \begin{align}
    C_2 &=   4 \int_0^1 dx \frac{1}{(-s-tx-i0)^2}
           \ +\ (s\leftrightarrow t) , \\
    C_1 &= - 4 \int_0^1 dx \frac{\ln(-s-tx-i0)}{(-s-tx-i0)^2}
           - 2 \int_0^1 d^2x \frac{1}{(-s-tx_1x_2-i0)^2}
           \ +\ (s\leftrightarrow t) , \\
\begin{split}
    C_0 &=   2 \int_0^1 dx \frac{\ln^2(-s-tx-i0)}{(-s-tx-i0)^2}
           + \int_0^1 d^2x \l4\{
               \frac{1}{x_1} \frac{-12\ln(1+x_1)-4\ln(1+x_1x_2)}{(-s-tx_2-i0)^2} \\
        &\hspace{50pt}
               + \frac{-4\ln(1+x_2)+2\ln(-s-tx_1x_2-i0)}{(-s-tx_1x_2-i0)^2}
             \r4\}
           \ +\ (s\leftrightarrow t) .
\end{split}
  \end{align}
\end{subequations}
The coefficients of Laurent expansion $C_2$, $C_1$ and $C_0$ are now
expressed in terms of multidimensional integrals, and in the Euclidean region
($s<0$ and $t<0$), they can be easily computed.

\section{Implementation of the sector decomposition on FORM}

In an intermediate stage of the sector decomposition for more complicated
loop integrals, one needs to handle a lot of terms and very large expression.
For example, the number of the generated sub-sectors
for the scalar massless on-shell triple box diagram is $\cal{O}(10000)$,
and then many more terms are produced by the $\ep$-expansion.
Most of popular computer algebra systems try to keep the expressions
in the physical memory.
When the expressions become large, the disk memory is used via
the virtual memory manager of the computer, and then
a computer algebra system can be extremely slowed down or does not work.
Therefore, in practice, there is a limit of the size of expressions
in such systems%
\footnote{%
  To avoid this limit, {\tt FIESTA}~\cite{Smirnov:2008py}
  uses a data base manager for storing the expressions on the disk efficiently.
}%
.
{\tt FORM} is designed to treat such huge expressions, which are larger
than the available physical memory, by using the disk with less penalties
in the performance, and therefore has strong advantage.

\begin{table}
  \begin{center}
    \begin{tabular}{|c|c|r|}
      \hline
      & Result and Error & Elapsed Time \\
      \hline
      $C_4$ & \ 0.{\bf{}20000000}661904E+01 \ $\pm$ \ 0.12485517866572E-05 & < 0.1s \\
      \hline
      $C_3$ &  -0.{\bf{}6000000}1069171E+01 \ $\pm$ \ 0.54096786712553E-05 & < 0.1s \\
      \hline
      $C_2$ &  -0.{\bf{}491674}20522766E+01 \ $\pm$ \ 0.17982248269768E-04 &   0.3s \\
      \hline
      $C_1$ & \ 0.{\bf{}11494738}132380E+02 \ $\pm$ \ 0.85267241358588E-04 &   9.5s \\
      \hline
      $C_0$ & \ 0.{\bf{}13801}183392483E+02 \ $\pm$ \ 0.29980115734943E-03 & 134.6s \\
      \hline
    \end{tabular}
    \caption{%
      The numerical result of the planar massless on-shell double box for
      $s=-1$ and $t=-1$ (CPU: Xeon 5160 3GHz). $C_i$ are the coefficients
      of $1/\ep^i$ (overall $\Gamma(3+2\ep)$ is excluded).
      With this parameter set, the integrand is relatively smooth (no strong peaks),
      and we used DCUHRE~\cite{berntsen-1991} as the multidimensional integrator.
      The goal of relative tolerance of the numerical integration is set as
      $E_\text{rel} = 10^{-6}$ for each integral.
      The boldfaced digits indicate they agree with the analytical values.
    }
    \label{tab:db-output}
  \end{center}
\end{table}

Our program uses {\tt FORM} for symbolic manipulations.
From the user input file describing the integral, {\tt FORM} produces
Fortran code for the integrand, after the sector decomposition.
The program has been checked by various multi-loop integrals, which
include planar and non-planar massless on-shell double boxes up to
$\cal{O}(\ep^0)$, massless three-loop propagators up to $\cal{O}(\ep^2)$,
etc. As an example, the numerical result of the planar massless on-shell
double box is shown in Table~\ref{tab:db-output}.

\section{Handling physical thresholds by the numerical extrapolation}

Consider the one-loop box diagram Eq.~\eqref{eq:box-integrand} again.
If either $s$ or $t$ is not negative and
the denominator
(or the argument of the logarithm)
of the integrand is not positive definite, it can become
to zero at some points in the integration region.
For such singularities arising from physical thresholds, the contour
deformation of Feynman parameters is used in~\cite{Nagy:2006xy}.
Actually the combination of the sector decomposition and the contour
deformation is applied in some practical
calculations~\cite{Lazopoulos:2007ix,Anastasiou:2007qb}.

Here we consider another possibility, the numerical extrapolation
method~\cite{deDoncker:2004fb}. In this method, we put $i\delta$ with
a small but finite $\delta$ instead of $i0$ in the denominator.
Then we can compute the integral $I(\delta)$ for a given $\delta$.
Calculating the sequence $\{I(\delta_k)\}$ for
$\delta_k = \delta_1 r^{k-1}$ ($k = 1,2,3,\ldots$, $0 < r < 1$) and
extrapolating them by an adequate method,
we can finally obtain the result of the integral in the limit of
$\delta \to 0$.

The numerical result of the massless one-loop on-shell box and
the massless one-loop box with two adjacent off-shell legs are shown in
Table~\ref{tab:0m-output} and~\ref{tab:2mh-output}, respectively.

\section{Summary}

The implementation of the sector decomposition algorithm on a symbolic
manipulation system {\tt FORM} is discussed. Thanks to {\tt FORM}'s
advantage for handling very large expressions, the limitation due to
the amount of available physical memory is greatly relaxed.
After isolating all IR divergences by the sector decomposition,
the numerical extrapolation is used for handling threshold singularities.
Several numerical results are also shown.

\acknowledgments

We would like to thank the members of MINAMI-TATEYA collaboration for discussions.

\begin{table}
  \begin{center}
    \begin{tabular}{|c|c|c|r|}
      \hline
      && Result and Error & Elapsed Time \\
      \hline
      $C_2$ & real      &  -0.{\bf{}16260162601}1506E-03 \ $\pm$ \ 0.216824649637068E-11 & < 0.1s \\
            & imaginary & \ 0.674151650287041{\bf{}E-15} \ $\pm$ \ 0.123815752782391E-14 &        \\
      \hline
      $C_1$ & real      & \ 0.{\bf{}98459363}6867288E-03 \ $\pm$ \ 0.972143597838532E-11 &   0.1s \\
            & imaginary &  -0.{\bf{}25541403}7543671E-03 \ $\pm$ \ 0.467152282813101E-11 &        \\
      \hline
      $C_0$ & real      &  -0.{\bf{}2388805}80536819E-02 \ $\pm$ \ 0.327822217718141E-10 &   0.8s \\
            & imaginary & \ 0.{\bf{}160867866}223496E-02 \ $\pm$ \ 0.159241557716135E-10 &        \\
      \hline
    \end{tabular}
    \caption{%
      The numerical result of the massless on-shell one-loop box for
      $s=123$ and $t=-200$ (CPU: Xeon 5160 3GHz). $C_i$ are
      the coefficients of $1/\ep^i$ (overall $\Gamma(2+\ep)$ is excluded).
      We used {\tt DQAGE} in ~\cite{piessens-1983}, which is
      one-dimensional integrator, iteratively. The goal of absolute and
      relative tolerance of the numerical integration are set as
      $E_\text{abs} = 10^{-15}$ and $E_\text{rel} = 10^{-8}$ for each integral.
      The boldfaced digits indicate they agree with the analytical values.
    }
    \label{tab:0m-output}
  \end{center}
\end{table}

\begin{table}
  \begin{center}
    \begin{tabular}{|c|c|c|r|}
      \hline
      && Result and Error & Elapsed Time \\
      \hline
      $C_2$ & real      &  -{\bf{}0.4065040650}28766E-04 \ $\pm$ \ 0.542061624092671E-12 &  < 0.1s \\
            & imaginary & \ 0.328759384083682{\bf{}E-15} \ $\pm$ \ 0.817570959150564E-15 &         \\
      \hline
      $C_1$ & real      & \ 0.{\bf{}3415630}70020333E-03 \ $\pm$ \ 0.288030340799012E-11 &    0.6s \\
            & imaginary & \ 0.{\bf{}127707018}334249E-03 \ $\pm$ \ 0.151057993697844E-11 &         \\
      \hline
      $C_0$ & real      &  -0.{\bf{}14929502}5193005E-02 \ $\pm$ \ 0.176182730120938E-10 & 1105.9s \\
            & imaginary &  -0.{\bf{}2874559}42345271E-03 \ $\pm$ \ 0.846787009393147E-10 &         \\
      \hline
    \end{tabular}
    \caption{%
      The numerical result of the massless one-loop box with two adjacent
      off-shell legs for $s=123$, $t=-200$, $p_3^2=50$ and $p_4^2=60$.
      Other conditions are same as Table~2. 
    }
    \label{tab:2mh-output}
  \end{center}
\end{table}

\end{document}